\documentclass[aps,prc,reprint,superscriptaddress]{revtex4-1}
\usepackage[colorlinks,linkcolor=blue,anchorcolor=blue,citecolor=blue,urlcolor=blue]{hyperref}
\usepackage{graphicx}
\usepackage{epstopdf}
\usepackage{dcolumn}
\usepackage{textcomp}
\usepackage{gensymb}
\usepackage{booktabs}
\usepackage{amsmath}
\usepackage{amssymb}
\usepackage{color}
\usepackage{hyperref}
\usepackage{url}
\usepackage{siunitx}
\usepackage{subfigure}
\usepackage{longtable}
\usepackage{array}
\usepackage{booktabs}
\usepackage{verbatim}
\newcolumntype{x}{D{.}{.}{6.6}}
\newcolumntype{y}{D{.}{.}{5.6}}
\newcolumntype{z}{D{.}{.}{5.7}}
\newcolumntype{f}{D{.}{.}{7.9}}
\newcolumntype{e}{D{.}{.}{6.6}}
\emergencystretch=\maxdimen
\begin{document}

%\preprint{APS/123-QED}

\title{Nuclear Moments of Germanium Isotopes around $N$ = 40}% Force line breaks with \\

\author{A. Kanellakopoulos}
 \affiliation{Instituut voor Kern- en Stalingsfysica, KU Leuven, B-3001, Leuven, Belgium}

\author{X. F. Yang}
 \email{xiaofei.yang@pku.edu.cn}
 \affiliation{School of Physics and State Key Laboratory of Nuclear Physics and Technology, Peking University, Beijing 100871, China}
 \affiliation{Instituut voor Kern- en Stalingsfysica, KU Leuven, B-3001, Leuven, Belgium}
 
\author{M. L. Bissell}
 \affiliation{Department of Physics and Astronomy, The University of Manchester, Manchester, M13 9PL, United Kingdom}

\author{\mbox{M. L. Reitsma}}
 \affiliation{Faculty of Science and Engineering, Van Swinderen Institute for Particle Physics and Gravity, University of Groningen, 9747 AG Groningen, The Netherlands}

\author{S. W. Bai}
 \affiliation{School of Physics and State Key Laboratory of Nuclear Physics and Technology, Peking University, Beijing 100871, China}

\author{J. Billowes}
 \affiliation{Department of Physics and Astronomy, The University of Manchester, Manchester, M13 9PL, United Kingdom}
 
\author{K. Blaum}
 \affiliation{Max-Planck-Institut f\"{u}r Kernphysik, D-69117 Heidelberg, Germany}
 
\author{\mbox{A. Borschevsky}}
 \affiliation{Faculty of Science and Engineering, Van Swinderen Institute for Particle Physics and Gravity, University of Groningen, 9747 AG Groningen, The Netherlands}
 
\author{\mbox{B. Cheal}}
 \affiliation{Oliver Lodge Laboratory, Oxford Street, University of Liverpool, Liverpool, L69 7ZE, United Kingdom}
 
\author{C. S. Devlin}
 \affiliation{Oliver Lodge Laboratory, Oxford Street, University of Liverpool, Liverpool, L69 7ZE, United Kingdom}

\author{R. F. Garcia Ruiz}
  \altaffiliation[Present address: ]{Massachusetts Institute of Technology, Cambridge, MA, USA}
  \affiliation{Experimental Physics Department, CERN, CH-1211 Geneva 23, Switzerland}
 
\author{H. Heylen}
 \affiliation{Experimental Physics Department, CERN, CH-1211 Geneva 23, Switzerland} 
 
\author{S. Kaufmann}
 \affiliation{Institut f\"{u}r Kernphysik, TU Darmstadt, D-64289 Darmstadt, Germany}
 \affiliation{Institut f\"{u}r Kernchemie, Universit\"{a}t Mainz, D-55128 Mainz, Germany}

\author{K. K\"{o}nig}
 \altaffiliation[Present adress: ]{National Superconducting Cyclotron Laboratory, Michigan State University, East Lansing, Michigan 48824, USA}
 \affiliation{Institut f\"{u}r Kernphysik, TU Darmstadt, D-64289 Darmstadt, Germany}

\author{\mbox{\'{A}. Koszor\'{u}s}}
 \altaffiliation[Present address: ]{Oliver Lodge Laboratory, Oxford Street, University of Liverpool, Liverpool, L69 7ZE, United Kingdom}
 \affiliation{Instituut voor Kern- en Stalingsfysica, KU Leuven, B-3001, Leuven, Belgium} 

\author{S. Lechner}
 \affiliation{Experimental Physics Department, CERN, CH-1211 Geneva 23, Switzerland}
 \affiliation{Technische Universit\"{a}t Wien, Karlsplatz 13, AT-1040 Wien, Austria}

\author{S. Malbrunot-Ettenauer}
 \affiliation{Experimental Physics Department, CERN, CH-1211 Geneva 23, Switzerland}
 
\author{R. Neugart}
 \affiliation{Max-Planck-Institut f\"{u}r Kernphysik, D-69117 Heidelberg, Germany}
 \affiliation{Institut f\"{u}r Kernchemie, Universit\"{a}t Mainz, D-55128 Mainz, Germany}
 
\author{G. Neyens}
 \affiliation{Instituut voor Kern- en Stalingsfysica, KU Leuven, B-3001, Leuven, Belgium}
 \affiliation{Experimental Physics Department, CERN, CH-1211 Geneva 23, Switzerland}

\author{W. N\"{o}rtersh\"{a}user} 
 \affiliation{Institut f\"{u}r Kernphysik, TU Darmstadt, D-64289 Darmstadt, Germany}
 
\author{\mbox{T. Ratajczyk}}
 \affiliation{Institut f\"{u}r Kernphysik, TU Darmstadt, D-64289 Darmstadt, Germany}

\author{L. V. Rodr\'{i}guez}
 \altaffiliation[Present address: ]{Experimental Physics Department, CERN, CH-1211 Geneva 23, Switzerland}
 \affiliation{Max-Planck-Institut f\"{u}r Kernphysik, D-69117 Heidelberg, Germany}
 \affiliation{Institut de Physique Nucl\'{e}aire, CNRS-IN2P3, Universit\'{e} Paris-Sud, Universit\'{e} Paris-Saclay, 91406 Orsay, France}
 
\author{S. Sels}
 \altaffiliation[Present address: ]{Instituut voor Kern- en Stalingsfysica, KU Leuven, B-3001, Leuven, Belgium}
 \affiliation{Experimental Physics Department, CERN, CH-1211 Geneva 23, Switzerland}
 
\author{S. J. Wang}
 \affiliation{School of Physics and State Key Laboratory of Nuclear Physics and Technology, Peking University, Beijing 100871, China}

\author{L. Xie}
 \affiliation{Department of Physics and Astronomy, The University of Manchester, Manchester, M13 9PL, United Kingdom}
 
\author{Z. Y. Xu}
 \altaffiliation[Present address: ]{Department of Physics and Astronomy, University of Tennessee, 37996 Knoxville, TN, USA}
 \affiliation{Instituut voor Kern- en Stalingsfysica, KU Leuven, B-3001, Leuven, Belgium}
 
\author{D. T. Yordanov}
  \affiliation{Institut de Physique Nucl\'{e}aire, CNRS-IN2P3, Universit\'{e} Paris-Sud, Universit\'{e} Paris-Saclay, 91406 Orsay, France}

\date{\today}% It is always \today, today,

\begin{abstract}
Collinear laser spectroscopy measurements were performed on $^{69,71,73}$Ge isotopes ($Z = 32$) at ISOLDE-CERN. The hyperfine structure of the $4s^2 4p^2 \, ^3P_1 \rightarrow 4s^2 4p 5s \, ^3P_1^o$ transition of the germanium atom was probed with laser light of 269 nm, produced by combining the frequency-mixing and frequency-doubling techniques. The hyperfine fields for both atomic levels were calculated using state-of-the-art atomic relativistic Fock-space coupled-cluster calculations. A new $^{73}$Ge quadrupole moment was determined from these calculations and previously measured precision hyperfine parameters, yielding $Q_{\rm s}$ = $-$0.198(4) b, in excellent agreement with the literature value from molecular calculations. The moments of $^{69}$Ge have been revised: \mbox{$\mu$ = +0.920(5) $\mu_{N}$} and $Q_{\rm s}$= +0.114(8) b, and those of $^{71}$Ge have been confirmed. The experimental moments around $N = 40$ are interpreted with large-scale shell-model calculations using the JUN45 interaction, revealing rather mixed wave function configurations, although their $g$-factors are lying close to the effective single-particle values. Through a comparison with neighboring isotones, the structural change from the single-particle nature of nickel to deformation in germanium is further investigated around $N = 40$.
\end{abstract}

\maketitle

\section{\label{sec:intro}Introduction}

Over the years, structural changes have been intensively investigated in the region between the semi-magic $^{68}$Ni and the doubly-magic $^{78}$Ni~\cite{Sorlin2002, 78NiPRL, Bissell2016, 78Ninature, Wraith2017, DeGroote2017}. The investigation involves multiple experimental methods as well as theoretical models, which provide various nuclear properties (masses, spins, life-times, transition probabilities, excitation energies, moments and radii), aiming to get a global view of the nuclear structure in this region. As studies deepen, more interesting nuclear phenomena appear. Some examples are: collective effects occur around and above $N = 40$~\cite{Xie2019, Cheal2010, Aoi2010, Chiara2011}; the proton $p_{3/2}$ and $f_{5/2}$ orbitals invert as neutrons are filling the neutron $g_{9/2}$ orbital due to the tensor part of the monopole interaction~\cite{Flanagan2009, Cheal2010, Wraith2017}; a spherical shape coexists with a deformed state around the neutron closed shells \mbox{$N = 40, 50$~\cite{CRIDER2016108, Yang2018, Yang2016}}; the indication of a weak sub-shell effect observed at $N = 40$ disappears quickly with more protons added above $Z = 28$~\cite{Bissell2016, Xie2019}.

\begin{figure*}[!t]
\includegraphics[width=0.9\textwidth]{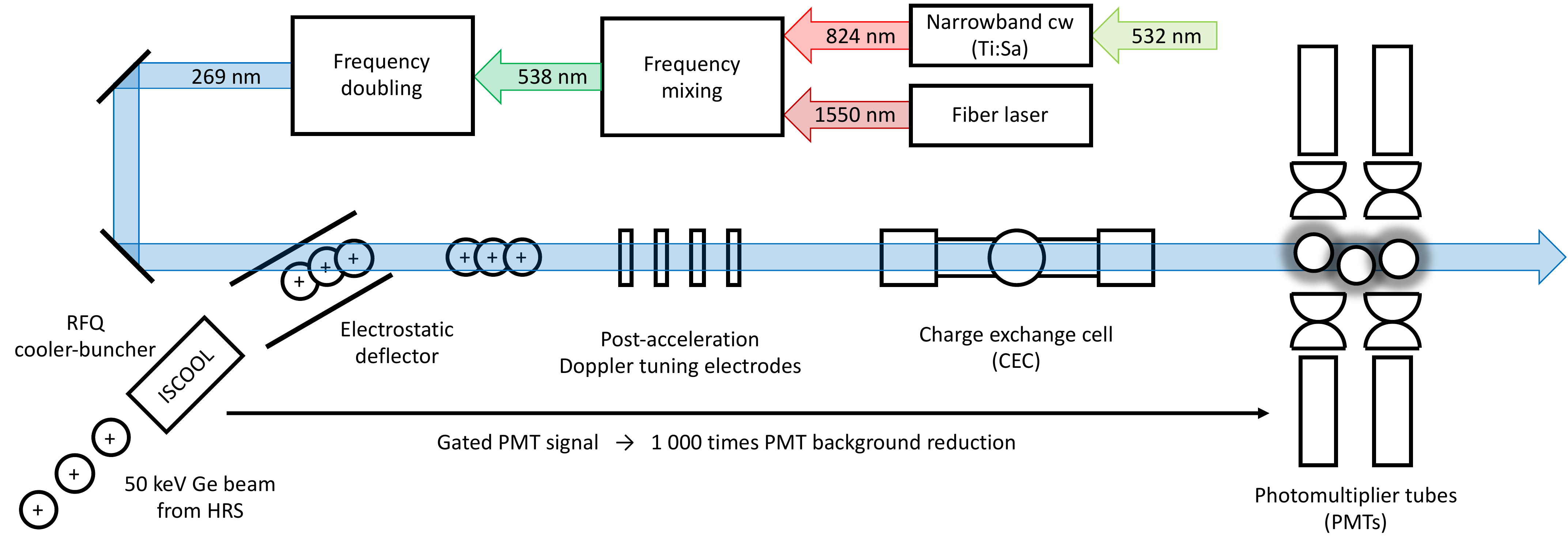}
\vspace{-4mm}
\caption{\label{fig:COLLAPS} A schematic view of the COLLAPS experimental setup and the laser frequency-mixing and frequency-doubling system. For details see text.}
\vspace{-5mm}
\end{figure*}

The nuclear properties of ground and isomeric states, such as spins, moments and charge radii, have made significant contributions in elucidating the above-mentioned nuclear phenomena. This was achieved by laser spectroscopy measurements on the isotopic chains of nickel ($Z =28$), copper ($Z =29$), zinc ($Z =30$) and gallium ($Z =31$)~\cite{Flanagan2009, Cheal2010, Vingerhoets2010, Procter2012, Bissell2016, DeGroote2017, Wraith2017, Xie2019, Simon2020, Yang2016, Yang2018, DeGroote2020}. Nuclear spins and magnetic dipole moments are sensitive probes of the single-particle nature or configuration mixing of the wave functions~\cite{Flanagan2009, Cheal2010, Wraith2017, DeGroote2017}, while the electric quadrupole moment and charge radii tell us more about the nuclear shape and collectivity~\cite{Yang2018, Bissell2016, Yang2016}.

Deformation, triaxiality and shape coexistence have been observed for zinc isotopes, which have also been reported for germanium isotopes around $N = 40$ based on $\gamma$-spectroscopy and reaction experiments~\cite{Xie2019, Ayangeakaa2016, Sun2014, Toh2013}. Germanium ($Z = 32$) has four protons outside the $Z = 28$ major shell closure, which may induce additional correlations leading to a complex wavefunction and collective effects for the low-lying states of germanium isotopes around $N = 40$~\cite{Ayangeakaa2016, Heyde2011, Heyde2016}. This collective feature can be further investigated with the nuclear moment measurements of germanium isotopes around $N = 40$ in combination with wave function calculations using large-scale shell-model interactions.

Until now, collinear laser spectroscopy has not been applied to study germanium isotopes, due to the fact that (1) germanium species cannot be easily produced at ISOL facilities, and (2) the suitable fine structure transitions are not easily accessible with standard laser equipment. In this article, we report the first hyperfine structure (hfs) measurement in the $4s^2 4p^2 \, ^3P_1 \rightarrow 4s^2 4p 5s \, ^3P_1^o$ transition of $^{69,71,73}$Ge atoms by combining the frequency-mixing and frequency-doubling laser techniques. State-of-the-art relativistic Fock-Space Coupled-Cluster (FSCC) calculations of the atomic hyperfine fields and electric field gradients have been performed, showing a good agreement with the experimental values, and thus benchmarking these atomic theories. The calculations also reveal that incorrect nuclear moments have been reported for $^{69}$Ge from earlier atomic beam magnetic resonance studies~\cite{Oluwole1970}. The extracted nuclear moments are compared with large-scale shell-model calculations using the JUN45 interaction in the $f_{5/2}pg_{9/2}$ model space in order to understand the collective effects. The systematic comparison of the nuclear moments of the 9/2$^{+}$ states in zinc ($Z = 30$), germanium ($Z = 32$) and selenium \mbox{($Z=34$)} allow the investigation of the structural evolution from single-particle to collective effects around $N = 40$ as more protons are added above the $Z = 28$ shell closure.

\section{Experimental procedure}

The experiment was performed at the collinear laser spectroscopy setup, COLLAPS~\cite{Neugart2017}, located at ISOLDE-CERN.
The radioactive germanium isotopes were produced by a 1.4-GeV proton beam impinging on a $\textrm{ZrO}_2$ target with a sulphur leak. The sulphur vapour was introduced into the target material where the sulphur atoms bind to form volatile GeS molecules.  These molecules easily diffuse out of the target and then disassociate under electron impact to form germanium ions inside a plasma source. The germanium ions were accelerated to 50\;keV, mass separated using the high-resolution isotope separator (HRS) and subsequently cooled and bunched in a gas-filled linear Paul trap (ISCOOL)~\cite{Mane2009}. The accumulated ions were released in short bunches with a typical temporal width of 5 $\mu$s every 5 ms. Due to the large isobaric contamination in all of the Ge beams \cite{Koster2003}, the accumulation time was optimized at 5~ms to avoid the overfilling of the ISCOOL.

\begin{table*}[!t]
\caption{\label{tab:AandB}%
Magnetic and electric hfs constants ($A$ and $B$) for both atomic states ($4s^2 4p^2 \, ^3P_1$  and $4s^2 4p 5s \, ^3P_1^o$) obtained from this work. The numbers from Refs.~\cite{Oluwole1970, Childs1963, Childs1966} are also summarized as a comparison.}
\renewcommand{\arraystretch}{1.2}
\begin{ruledtabular}
\begin{tabular}{cccccccccc}
$A$ & $N$ & $I^\pi$ & $T_{1/2}$ & $A_{\rm l}^{\rm lit}$ (MHz) & $A_{\rm l}$ (MHz) & $A_{\rm u}$ (MHz) & $B_{\rm l}^{\rm lit}$ (MHz) & $B_{\rm l}$ (MHz) & $B_{\rm u}$ (MHz) \\
\colrule
69 & 37 & 5/2$^-$ & 39.05(10) h & $\mp$23.40(3)~\cite{Oluwole1970} & $-$29.0(7) & $+$494.6(9) & $\pm$8.28(8)~\cite{Oluwole1970} & $+$32(2)\footnotemark[1] & $-$21(3)\footnotemark[1] \\
71 & 39 & 1/2$^-$ & 11.43(3) d & $-$87.005(3)~\cite{Childs1963} &  & $+$1460.8(16) &  &  &  \\
73 & 41 & 9/2$^+$ & stable & $+$15.5480(18)~\cite{Childs1966} &  & $-$262.2(12) &  $-$54.566(9)~\cite{Childs1966} & & $+$40(6)
\footnotetext[1]{Correlation C($B_{\rm l}$, $B_{\rm u}$) = -0.445}
\end{tabular}
\end{ruledtabular}
\end{table*}

As shown in \mbox{Fig.~\ref{fig:COLLAPS}}, the bunched ion beam was then deflected into the COLLAPS beamline, where it was neutralized in-flight by passing through a charge exchange (CE) cell filled with sodium vapor. The state $4s^2 4p^2 \, ^3P_1$ of the germanium atom was populated in the neutralization process \cite{Vernon2019}. A frequency fixed continuous-wave (cw) laser beam, overlapped with the germanium beam in a collinear geometry, was used to probe the $4s^2 4p^2 \, ^3P_1 - 4s^2 4p 5s \, ^3P_1^o$ \mbox{(269.13411 nm)} transition of the germanium atom. By applying a varying voltage to the germanium ions before entering the CEC, neutralized germanium atoms were resonantly excited to the $4s^2 4p 5s \, ^3P_1^o$ state through Doppler tuning. The emitted fluorescence photons from the resonantly excited atoms were detected as a function of the tuning voltage, by the use of 4 photomultiplier tubes (PMTs) \cite{Kreim2014}. A software time gate was applied in order to select photons only when each bunch of germanium atoms passes and de-excites in front of the PMTs. This resulted in a reduction of the background from laser stray light, non-resonantly scattered photons and PMT dark counts by a factor of about $10^3$.

\begin{figure}[!t]
\includegraphics[width=0.5\textwidth]{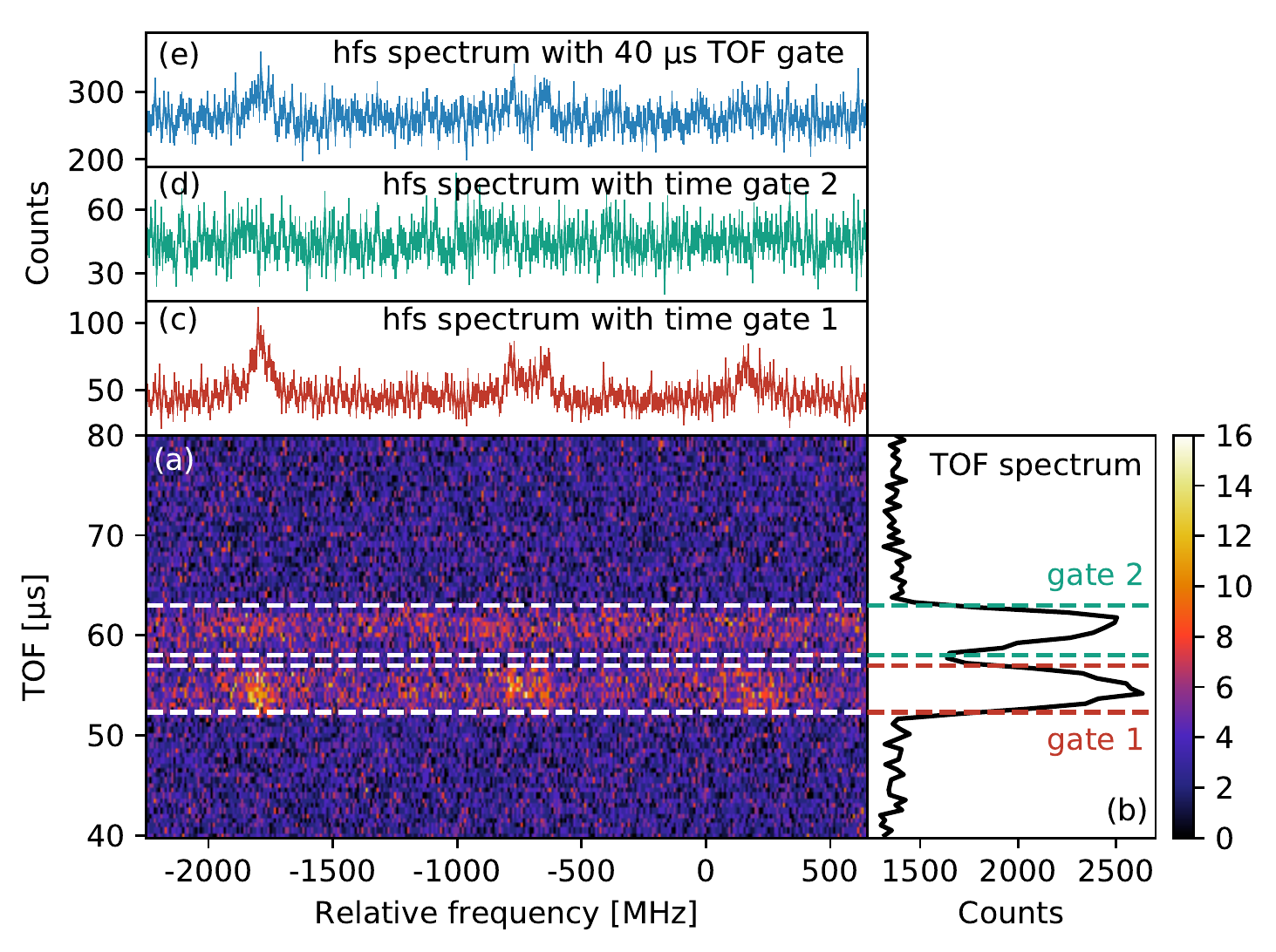}
\vspace{-4mm}
\caption{\label{fig:HFSandTOFheatmap}Color-coded 2D plot with time vs frequency (voltage step) vs counts acquired for $^{73}$Ge with the new DAQ system (TILDA) of COLLAPS. The two components presented in the TOF spectrum are corresponding to the germanium atom bunch (gate 1 in red) and the possible molecular contamination bunch (gate 2 in green). Gating on the germanium atom bunch allows the hfs spectrum to be reconstructed with a significantly reduced background.}
\vspace{-4mm}
\end{figure}

To take full advantage of ISCOOL and to retain as much information on the beam structure in the recorded data as possible, a new data acquisition system (DAQ) was introduced at COLLAPS. It was developed at the TRIGA-SPEC setup in Mainz \cite{Kaufmann.2015} and here we report its online-operation. The main advantage of this DAQ is the time-resolved photon detection. In order to reduce the photon background, the detection of photons has to be restricted to the time window when the bunched beam is traversing the detection region. Previously, this was realized by hardware gating, which had to be adapted in advance for each isotope according to the time-of-flight (TOF) recorded. With the new DAQ, photons are recorded with a time stamp relative to the ion extraction trigger of ISCOOL. This is shown for $^{73}$Ge in Fig.~\ref{fig:HFSandTOFheatmap}(a): the $x$-axis represents the scanning voltage, converted into frequency, while the $y$-axis is the TOF since the ISCOOL extraction pulse. The number of photons detected within a 500-ns interval during $n$-time extractions from ISCOOL are color-coded. These numbers are integrated along a specified time in Fig.~\ref{fig:HFSandTOFheatmap}(b), which provides the time structure of the  bunch. Integration along a fixed frequency reveals the hfs resonance spectrum of the isotope, as illustrated in Fig.~\ref{fig:HFSandTOFheatmap}(c). The spectra in \mbox{Fig.~\ref{fig:HFSandTOFheatmap}(c)-(d)} are obtained by selecting events within the specific time window indicated with dashed horizontal lines in Fig.~\ref{fig:HFSandTOFheatmap}(a) and with their corresponding colored dashed lines in Fig.~\ref{fig:HFSandTOFheatmap}(b).

\begin{figure}[!t]
\includegraphics[width=0.5\textwidth]{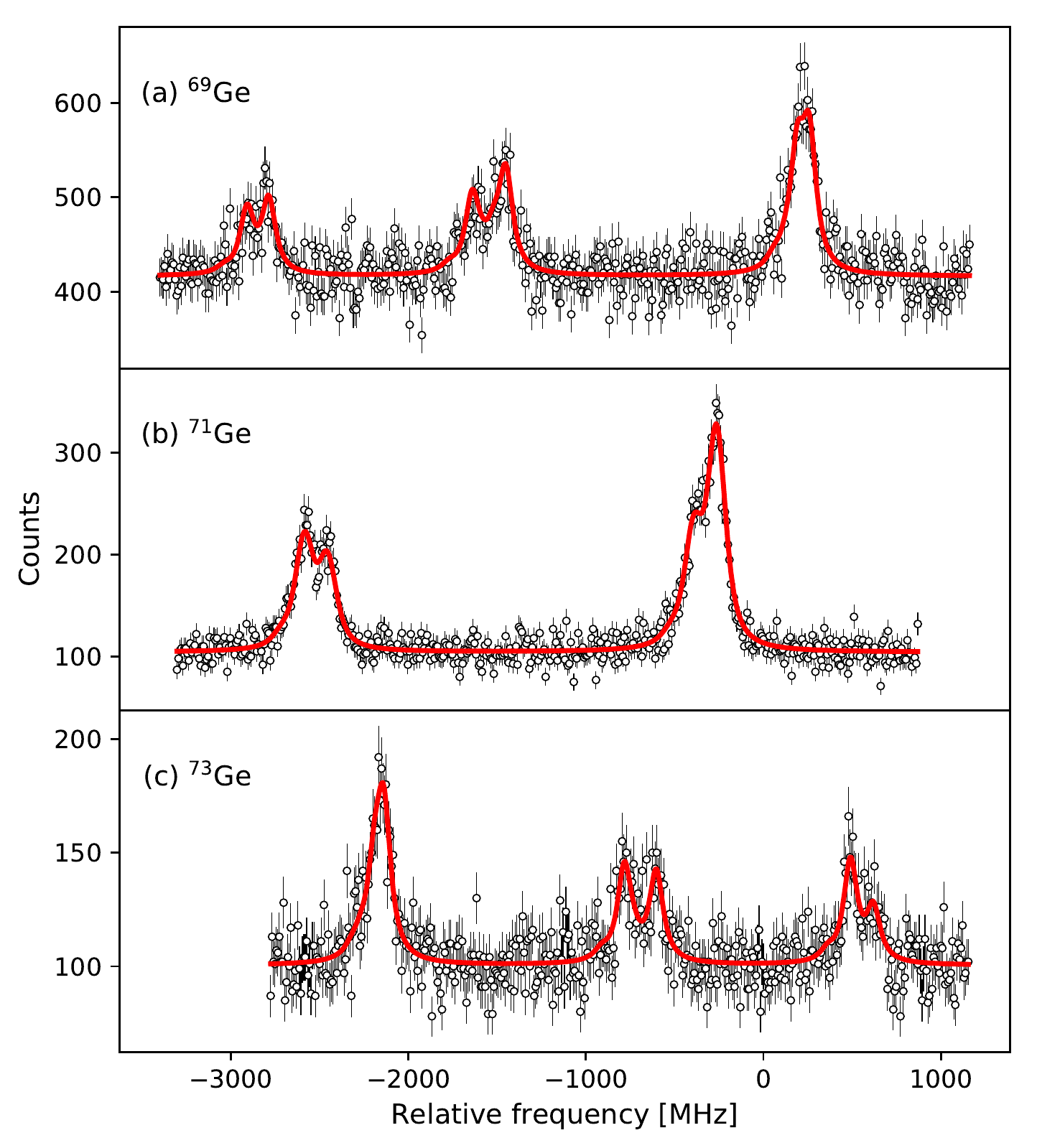}
\vspace{-4mm}
\caption{\label{fig:hfs}Hyperfine structure spectra of $^{69,71,73}$Ge in the $4s^2 4p^2 \, ^3P_1 \rightarrow 4s^2 4p 5s \, ^3P_1^o$ transition. The red lines show the best fit with a Voigt line profile. Note that the production of the known short-lived isomers of $^{71,73}$Ge (T$_{1/2}$($^{71m}$Ge) = 20.41(18)~ms and T$_{1/2}$($^{73m}$Ge) = 499(11)~ms) \cite{Abusaleem2011, Singh2019} is at least 100 times lower than that of the ground state \cite{Bissell:2157172}. Thus, these isomers could not be observed in this experiment.}
\vspace{-4mm}
\end{figure}

In Fig.~\ref{fig:HFSandTOFheatmap}(a) three regions can be clearly distinguished. Outside of the dashed lines, the background is dominated by stray laser light, which is weak compared to the more intense signal within the time gate of the bunch. One can clearly see that the bunch is separated into two parts, the first one arriving after about 52\,$\mu$s, while the second one is delayed by another 5\,$\mu$s. The time interval between two bunches released from ISCOOL is about 1000 times longer. Thus, the two consecutive \lq bunches' seen in Fig.~\ref{fig:HFSandTOFheatmap}(a) are from photons emitted from two different species. Those with a heavier mass arrive later in front of the PMT's. In the first time window (gate 1) clear resonance spots are seen, as well as an additional beam-related but frequency-independent background (see also Fig.~\ref{fig:HFSandTOFheatmap}(c)). This background is attributed to the de-excitation of all contaminant neutral particles that were excited in the CE process. The second part of the bunch (gate 2) does not exhibit any resonance-like structures, only beam-related photon background, as seen in Fig.~\ref{fig:HFSandTOFheatmap}(d). It is assumed that this beam-related photon background is due to laser-light scattering from the molecular contamination in the beam and de-excitations of these molecules excited via the CE process.

With the latest DAQ, these individual parts of the bunch can be clearly separated in online and offline analysis. Therefore, by gating on the first component (gate 1) instead of the plotted range (40\,$\mu$s) of the TOF spectrum, the hfs spectrum of $^{73}$Ge was obtained with a much improved signal-to-background ratio, as shown in Fig.~\ref{fig:HFSandTOFheatmap}(c) and Fig.~\ref{fig:HFSandTOFheatmap}(e), respectively.

One challenge for the laser spectroscopy measurement of germanium atoms is the production of the 269 nm cw laser light. It requires frequency-doubling of the wavelength 538 nm. This wavelength is lying in the \lq green gap' region, which is not covered by the commonly used continuous wave (cw) Ti:Sa and dye laser systems. To bridge this wavelength gap, a frequency mixing method was employed. As shown in \mbox{Fig.~\ref{fig:COLLAPS}}, a 824 nm laser beam from a Ti:Sa laser cavity (Sirah Matisse-2) and a 1550 nm laser beam generated by a single-frequency fiber laser (Koheras Boostik E 15) are superimposed and single-pass through a periodically poled crystal in a frequency mixing unit (Sirah MixTrain) to generate the sum frequency of 538 nm ($\frac{1}{\lambda}=\frac{1}{\lambda_1}+\frac{1}{\lambda_2}$). This light is then coupled into a frequency doubling unit (Sirah WaveTrain) to produce 269 nm light with an output power around 90 mW. A small fraction of the output light from the frequency mixing unit was locked to a wavelength meter (HighFinesse WSU10), which was regularly calibrated with a stabilized diode laser (Toptica DLPRO780) locked to the $F = 2  \rightarrow F = 3$ hyperfine transition of the D1 line in $^{87}$Rb.

\section{Results}

\begin{figure}[!t]
\includegraphics[width=0.5\textwidth]{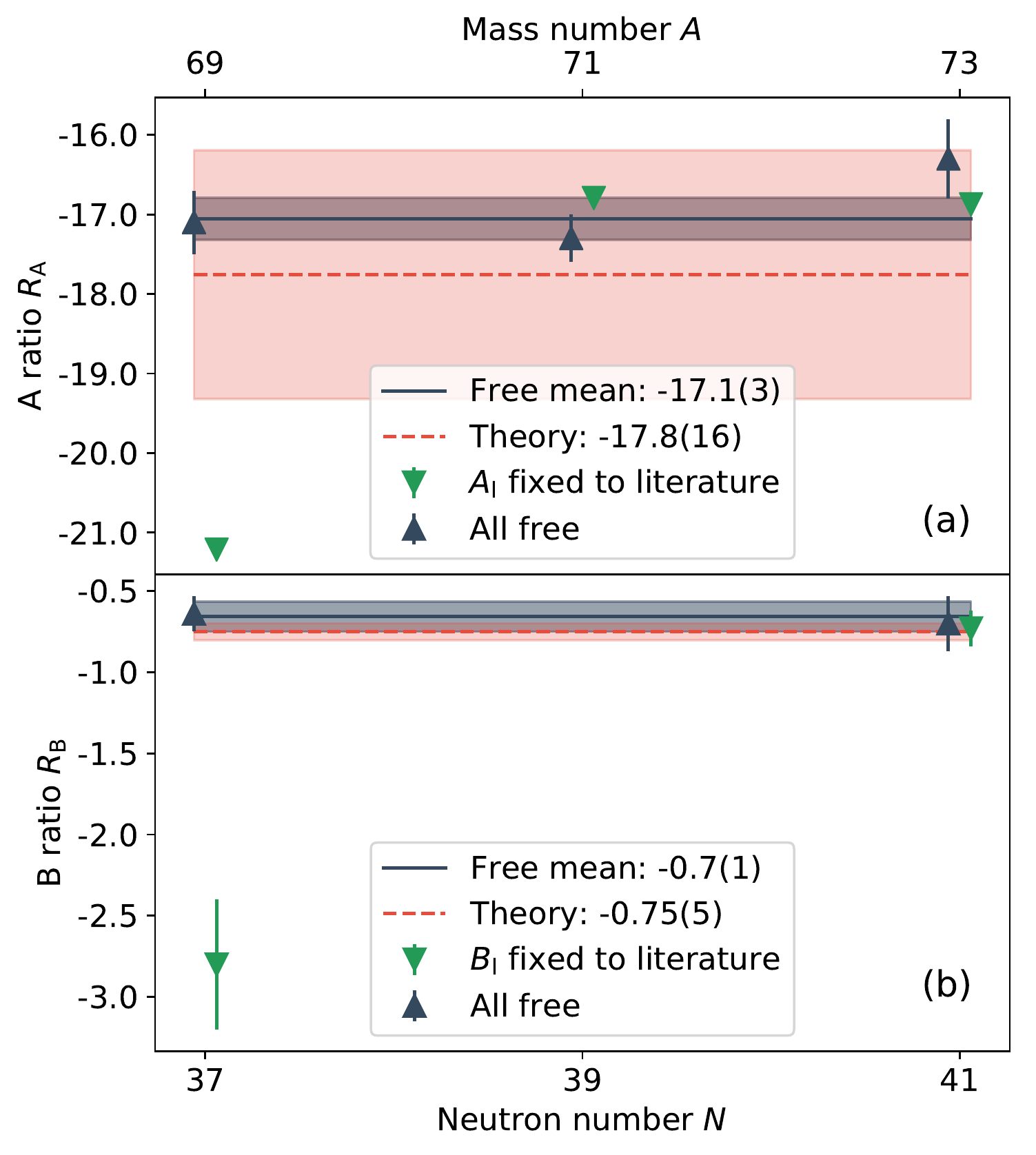}
\vspace{-5mm}
\caption{\label{fig:AandB} Ratio of the $A$ (a) and $B$ hyperfine constants (b) for the $4p^2 \, ^3P_1$ (lower) and $4p5s \, ^3P_1^o$ (upper) atomic states, deduced with two fitting procedures. One with the $A_{\rm l}$ and $B_{\rm l}$ values for each isotope fixed (green) to the high-precision value reported in literature \cite{Oluwole1970, Childs1963, Childs1970}, and one with all hyperfine constants as free fit parameters (black). The hyperfine constants ratios calculated with atomic relativistic FSCC are shown by red dashed lines.}
\vspace{-5mm}
\end{figure}

The obtained hfs spectra of $^{69,71,73}$Ge isotopes, as shown in Fig.~\ref{fig:hfs}, were fitted with a $\chi^2$-minimization Python routine using the SATLAS package~\cite{Gins2018}. A Voigt profile with one side peak was used to compensate for the slightly asymmetric resonance peaks, which resulted from the energy loss due to the population of higher atomic states or the collision-excitation in the charge exchange process~\cite{Neugart1977, Bendali1986}. All the recorded hfs spectra of each individual isotope were fitted simultaneously, generating one optimal reduced $\chi^2$, with magnetic dipole and electric quadrupole hyperfine constants ($A$ and $B$) as common fit parameters for each isotope.

\begin{table}[!b]
\caption{\label{tab:HFSratios}%
Deduced $R_{\rm A} = A_{\rm u} / A_{\rm l}$ and $R_{\rm B} = B_{\rm u} / B_{\rm l}$ for $^{69,71,73}$Ge isotopes
with the $A_{\rm l}$ and $B_{\rm l}$ fixed to literature values or with all hyperfine constants as free fit parameters.}
\renewcommand{\arraystretch}{1.2}
\begin{ruledtabular}
\begin{tabular}{c|cc|cc}
 & \multicolumn{2}{c|}{$R_{\rm A}$} & \multicolumn{2}{c}{$R_{\rm B}$} \\
 & All free & $A_{\rm l}$ fixed to lit. & All free & $B_{\rm l}$ fixed to lit. \\
\colrule
$^{69}$Ge & $-$17.1(4) & $-$21.21(4) & $-$0.64(11) & $-$2.8(4) \\
$^{71}$Ge & $-$17.3(3) & $-$16.79(2) & & \\
$^{73}$Ge & $-$16.3(5) & $-$16.87(8) & $-$0.70(17) & $-$0.73(11)
\end{tabular}
\end{ruledtabular}
\end{table}

The $A$ and $B$ parameters for the lower atomic state ($4s^2 4p^2 \, ^3P_1$),  named $A_{\rm l}$ and $B_{\rm l}$ in the following, are reported in literature with high precision for each of the three \mbox{odd-$A$} germanium isotopes from atomic beam magnetic resonance experiments~\cite{Childs1963, Childs1966, Oluwole1970, Childs1970}, as summarized in \mbox{Tab.~\ref{tab:AandB}}. We therefore fitted our data in two ways: (1) with the $A_{\rm l}$ and $B_{\rm l}$ values fixed to the literature value and upper hyperfine constants as free fit parameters for each isotope, and (2) with all lower and upper hyperfine constants as free fit parameters. The ratio of the hyperfine parameters of the upper and lower atomic states ($R_{\rm A}$ = $A_{\rm u}$/$A_{\rm l}$, $R_{\rm B}$ = $B_{\rm u}$/$B_{\rm l}$) reflect the ratio of the atomic hyperfine fields of both levels. These ratios should be the same for all isotopes, apart from a possible small hyperfine anomaly that can cause some scatter of the order of at most a few percent in $R_{\rm A}$. In \mbox{Table~\ref{tab:HFSratios}}, as well as in \mbox{Fig.~\ref{fig:AandB}}, the ratio of $R_{\rm A}$ and $R_{\rm B}$ resulting from the two fit procedures are shown. When using the first fitting procedure, both the $A$ and $B$ factor ratios for $^{69}$Ge deviate strongly from those observed for the other isotopes, suggesting a problem with the reported hyperfine parameters of $^{69}$Ge \cite{Oluwole1970}, from which the literature nuclear moments have been deduced. When all four hyperfine parameters are free fit parameters, the deduced ratios of $R_{\rm A}$ and $R_{\rm B}$ are consistent for all three isotopes, as shown in \mbox{Tab.~\ref{tab:HFSratios}} and \mbox{Fig.~\ref{fig:AandB}}. To corroborate these conclusions, state-of-the-art atomic calculations were performed, as discussed in the following section.

\begin{table*}[!t]
\caption{\label{tab:AtomicTheory}%
Atomic $A_0$ and $q$ parameters extracted from $^{73}$Ge experimental results and calculated with atomic relativistic FSCC method.}
\renewcommand{\arraystretch}{1.2}
\begin{ruledtabular}
\begin{tabular}{ccccccc}
At. state & $A^{\rm exp}$ (MHz) & $A^{\rm exp}_{\rm 0}$ (MHz) & $A^{\rm th}_{\rm 0}$ (MHz) & $B^{\rm exp}$ (MHz) & $q^{\rm exp}$ (MHz/b) & $q^{\rm th}$ (MHz/b) \\
\colrule
$4p^2 \, ^3P_1$ & $+$15.5480(18) \cite{Childs1966} & $-$79.667(10) & $-$74(6) & $-$54.566(9) \cite{Childs1966} & $+$278.4(14) & $+$277(8) \\
$4p^2 \, ^3P_2$ & $-$64.4270(7) \cite{Childs1966} & $+$330.12(2) & $+$321(11) & $+$111.825(13) \cite{Childs1966} & $-$571(3) & $-$564(17) \\
$4p5s \, ^3P_1^o$ & $-$262.2(12)\footnotemark[1] & $+$1343(6) & $+$1314(45) & $+$40(6)\footnotemark[1] & $-$204(31) & $-$208(13) \\
\end{tabular}
\footnotetext[1]{Hyperfine parameters measured in this work.}
\end{ruledtabular}
\end{table*}

\begin{table*}[!t]
\caption{\label{tab:MandQ}%
Magnetic dipole and electric quadrupole moments of $^{69, 71, 73}$Ge isotopes compared with JUN45. The numbers from Refs.\cite{Oluwole1970, Childs1966, Makulski2006, Kello1999} are also summarized as a comparison.}
\renewcommand{\arraystretch}{1.2}
\begin{ruledtabular}
\begin{tabular}{ccccccccc}
$A$ & $N$ & $I^\pi$ & $\mu^{\rm lit}$ ($\mu_N$) & $\mu^{\rm exp}$ ($\mu_N$) & $\mu^{\rm JUN45}$ ($\mu_N$) & $Q_{\rm s}^{\rm lit}$ (b) & $Q_{\rm s}^{\rm exp}$ (b) & $Q_{\rm s}^{\rm JUN45}$ (b) \\
\colrule
69 & 37 & 5/2$^-$ & 0.735(7)\cite{Oluwole1970} & $+$0.920(5) & $+$1.048 & $+$0.027(5)\cite{Oluwole1970} & $+$0.114(7) & $+$0.150 \\
71 & 39 & 1/2$^-$ & $+$0.54606(7) \cite{Childs1966} & $+$0.547(5) & $+$0.438 &  &  &  \\
73 & 41 & 9/2$^+$ & $-$0.87824(5) \cite{Makulski2006} & -0.904(21)\footnotemark[1]  & $-$0.955 & $-$0.196(1) \cite{Kello1999} & -0.198(4)\footnotemark[1] & $-$0.258 \\
\end{tabular}
\footnotetext[1]{Weighted mean of the moments extracted from the measured hyperfine parameters and calculated hyperfine fields of the 3 transitions in \mbox{Tab.~\ref{tab:AtomicTheory}}.}
\end{ruledtabular}
\end{table*}

\subsection{Atomic hyperfine field calculations }
To further investigate the aforementioned discrepancy between the hyperfine constants from literature and our experimental results for $^{69}$Ge, we performed atomic calculations to obtain the hyperfine fields of three atomic fine structure levels ($4s^24p^2$ $^3P_1$, $4s^24p^2$ $^3P_2$ and $4s^24p5s$ $^3P_1^o$), for which experimental information is available for the stable $^{73}$Ge isotope~\cite{Childs1966}.

High-quality treatment of relativistic and electron correlation effects is required to obtain accurate and reliable computational results. Therefore, the multireference relativistic FSCC method \cite{Kaldor1998, Visscher2001} is employed for the investigation of the electric field gradients (EFGs) ($q^{\rm th}$) and the hyperfine magnetic field constants ($A_0^{\rm th}$). This method was shown to be one of the most powerful approaches for treatment of spectra and properties of heavy many-electron atoms \cite{ABcoupledcluster}.

The FSCC method requires a closed shell reference state from which the ground state and excited states can be reached by adding or removing electrons. Neutral germanium has an open shell ground state electron configuration [Ar]$3d^{10}4s^24p^2$ and thus the closed shell Ge$^{2+}$ system ([Ar]$3d^{10}4s^2$) is used as the reference state. The ground state and excited states of interest are reached by adding two electrons to the corresponding virtual orbitals, which comprise the model space. The intermediate Hamiltonian (IH) approach is applied to avoid the intruder-state problem \cite{Landau2001}. The finite field method \cite{Pople1968} is used to determine the $q^{\rm th}$ and $A_0^{\rm th}$ properties, as described in Refs. \cite{Yakobi2007, Haase2020}. All calculations are carried out in the framework of the Dirac-Coulomb Hamiltonian and the nuclear charge distribution is modeled by a Gaussian function as described in Ref. \cite{Visscher1997}. The $q^{\rm th}$ calculations were carried out using the DIRAC15 program package \cite{DIRAC15}, while DIRAC17 was used for the $A_0^{\rm th}$ calculations \cite{DIRAC17}. The final values are obtained using the full 4-component DC Hamiltonian and the relativistic core-valence 4-zeta basis set of Dyall \cite{Dyall2006} (cv4z), augmented by three diffuse functions in each symmetry in an even-tempered fashion. A large model space was used, consisting of the $4p$ $5s$ ($4d$ $5p$ $6s$ $4f$ $5d$ $6p$ $7s$ $5f$ $5g$ $7p$ $6d$ $7d$ $8p$ $6g$ $8s$ $6f$) orbitals, where the orbitals in parentheses are in the intermediate space $P_i$. All the electrons were correlated and virtual orbitals with energies up to 500 a.u. were included in the calculation. The finite field perturbation strength $\lambda$ was $\num{1e-6}$ for the $q^{\rm th}$ calculations and $\num{1e-4}$ for $A_0^{\rm th}$. 

To estimate the uncertainty on the calculated values, we have investigated the effect of various computational parameters: the error from the limited size of the basis set, model space and virtual space, while also estimating the contribution from higher order excitations and the relativistic Gaunt term. These sources of error are combined by assuming them to be independent to give a total conservative uncertainty estimate for each calculated property. The uncertainty of $q^{\rm th}$ is 3.0 \% for the $4p^2$ ${}^3P_1$ and $4p^2$ states and 6.4 \% for $4p5s$ ${}^3P_1^o$, while for $A_0^{\rm th}$ the uncertainties are 7.8 \%, 3.5 \% and 3.4 \% respectively for these three states. The final recommended values and uncertainties for $q^{\rm th}$ and $A_0^{\rm th}$ are shown in \mbox{Tab.~\ref{tab:AtomicTheory}}. Further details on the procedure used for the calculation of these properties and the uncertainties can be found in Ref. \cite{Gustafsson2020}.

The calculated atomic observables, $A_0$ and $q$, are proportional to the magnetic hyperfine field ($B_0$) created by the electrons at the point of the nucleus, \mbox{$A_0 = B_0 / J$}, and to the electric field gradient ($V_{zz}$) created by the electronic cloud, \mbox{$q = e V_{zz}$}, respectively. Here, $J$ is the atomic spin. These atomic observables are related to the measured $A$ and $B$ parameters as follows: \mbox{$A_0 = A I / \mu$} and \mbox{$q = B / \, Q_{\rm s}$}, with $\mu$ and $Q_{\rm s}$ the nuclear magnetic dipole moment and electric quadrupole moment, respectively, and $I$ the nuclear spin. From the experimental hyperfine constants of the three atomic states taken from Refs. \cite{Childs1966} and measured in this work, as well as the experimental magnetic and quadrupole moments from \mbox{Refs.~\cite{Kello1999, Makulski2006}}, we can calculate the experimental $A_0$ and $q$ parameters for $^{73}$Ge. These can be directly compared with atomic theory calculations, as presented in \mbox{Tab.~\ref{tab:AtomicTheory}}. A remarkable agreement is achieved for both $A_0$ and $q$, for all three states within less than 5\%. We further compared the ratios ($R_{\rm A}$, $R_{\rm B}$) of hyperfine constants of two atomic states ($4p^2 \, ^3P_1$ and $4p5s \, ^3P_1^o$) measured in this work with the atomic calculation, as shown in \mbox{Fig.~\ref{fig:AandB}}. The ratios of the hyperfine constants from theory show a good agreement with experimental values for all three isotopes of $^{69,71,73}$Ge obtained with fitting procedure (2). This further supports our experimental result for the $A_{\rm l}$ and $B_{\rm l}$ hyperfine constants of $^{69}$Ge.

\subsection{Hfs constants and nuclear moments}
As mentioned above, both our experimental results and the atomic calculations are inconsistent with the literature hyperfine constants for the $4p^2 \, ^3P_1$ atomic state of $^{69}$Ge. A revision of the lower state hfs parameters of $^{69}$Ge is therefore suggested. From our measurements, the upper-state hyperfine parameters, $A_{\rm u}$ and $B_{\rm u}$, can also be determined for the first time for all isotopes, yielding an independent measurement of the nuclear moments. For $^{71,73}$Ge, the final $A_{\rm u}$ and $B_{\rm u}$ are obtained with $A_{\rm l}$ and $B_{\rm l}$ fixed to the literature values~\cite{Childs1963, Childs1966}, since they are known with high-precision. In \mbox{Tab.~\ref{tab:AandB}}, the revised hyperfine parameters of the lower atomic state for $^{69}$Ge, and the newly measured hyperfine constants of the upper atomic state for $^{69,71,73}$Ge are shown together with literature values.

A precise nuclear magnetic moment of $^{73}$Ge has recently been redetermined using gas-phase NMR measurements on GeH$_4$~\cite{Makulski2006} and a precise electric quadrupole moment was extracted from molecular microwave data of GeO and GeS~\cite{Kello1999}. Thus, the magnetic dipole $\mu$ and electric quadrupole $Q_{\rm s}$ moments of $^{69,71}$Ge can be calculated from the experimental $A$ and $B$ parameters in a model independent way relative to those of $^{73}$Ge, by using: 
\begin{equation}
\mu= \frac{IA}{I_{\rm ref}A_{\rm ref }} \mu_{\rm{ref}}
\label{EQ1}
\end{equation}
\vspace{-3mm}
\begin{equation}
Q_{\textrm{s}}= \frac{B}{B_{\rm ref}} Q_{\rm s,ref}
\label{EQ2}
\end{equation}

The large $A_{\rm u}$ hyperfine parameter is used to extract the magnetic moments of $^{69,71}$Ge via \mbox{Eq.~(\ref{EQ1})}. Since the hyperfine $B$ parameters for both atomic states are comparable in magnitude, two sets of electric quadrupole moments are extracted from $B_{\rm l}$ and $B_{\rm u}$ using \mbox{Eq.~(\ref{EQ2})}. We use the weighted average of the two quadrupole moments as the final value, after taking into account the correlation between the two hyperfine $B$ parameters \textcolor{red}{\cite{Leo:1987kd}}. The results are summarized in \mbox{Tab.~\ref{tab:MandQ}} together with the literature values.

For the reference isotope $^{73}$Ge, we also determine an independent set of nuclear moments, using the measured hyperfine parameters and calculated hyperfine fields for each of the three atomic states shown in \mbox{Tab.~\ref{tab:AtomicTheory}}. These moments are fully consistent with the literature values (\mbox{Tab.~\ref{tab:MandQ}}), in particular for the quadrupole moment reaching almost similar precision. This illustrates the significant progress that has been made in atomic calculations in the last few years.

\begin{figure}[!b]
\includegraphics[width=0.45\textwidth]{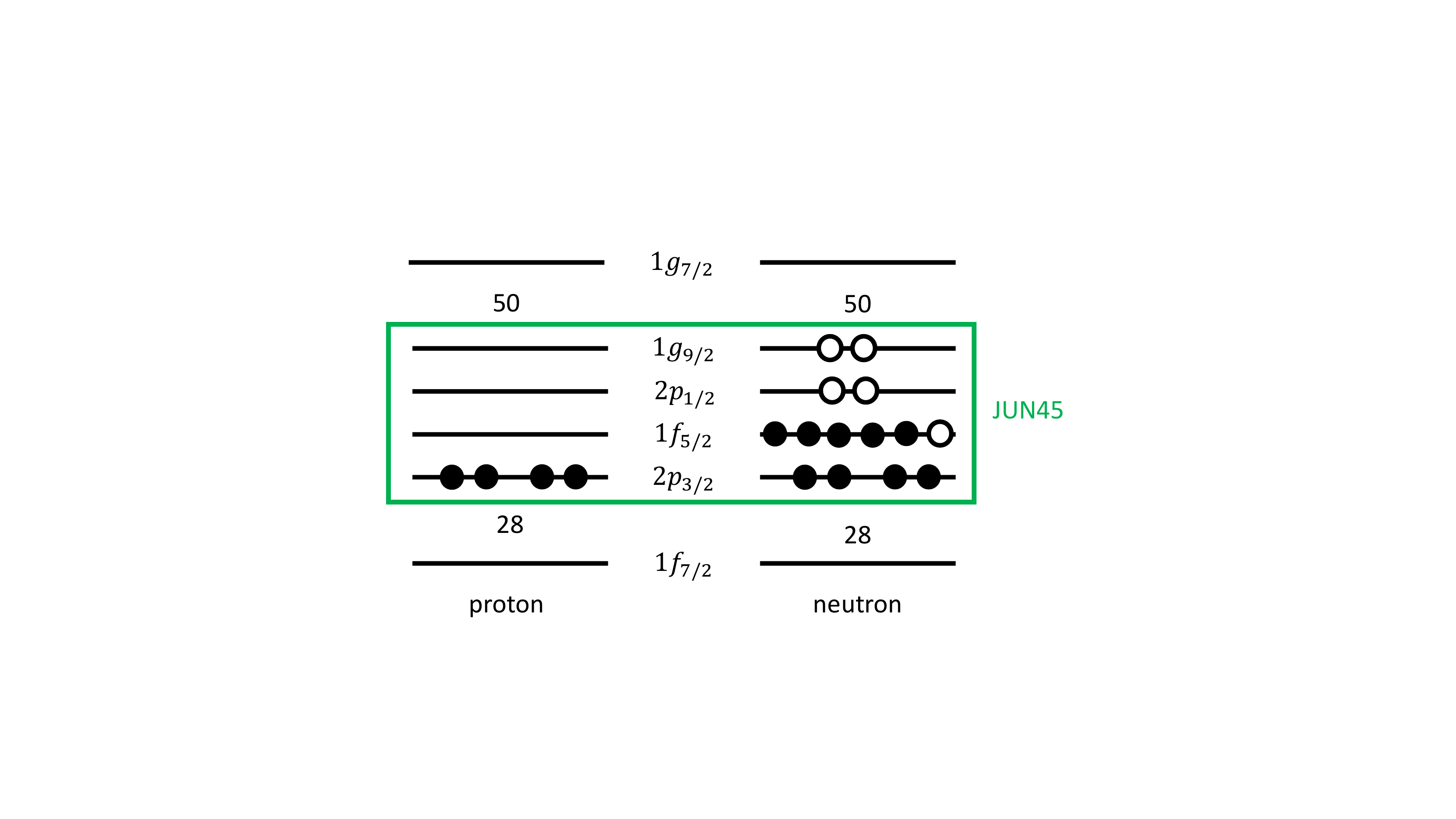}
\vspace{-3mm}
\caption{\label{fig:shells}Proton and neutron orbitals around the $Z, N$ = 28 and the $Z, N$ = 50 major shell closures according the nuclear shell-model. The model space of the effective interaction JUN45 is also marked.}
\end{figure}

\section{Discussion}

\begin{figure*}[!t]
\includegraphics[width=0.98\textwidth]{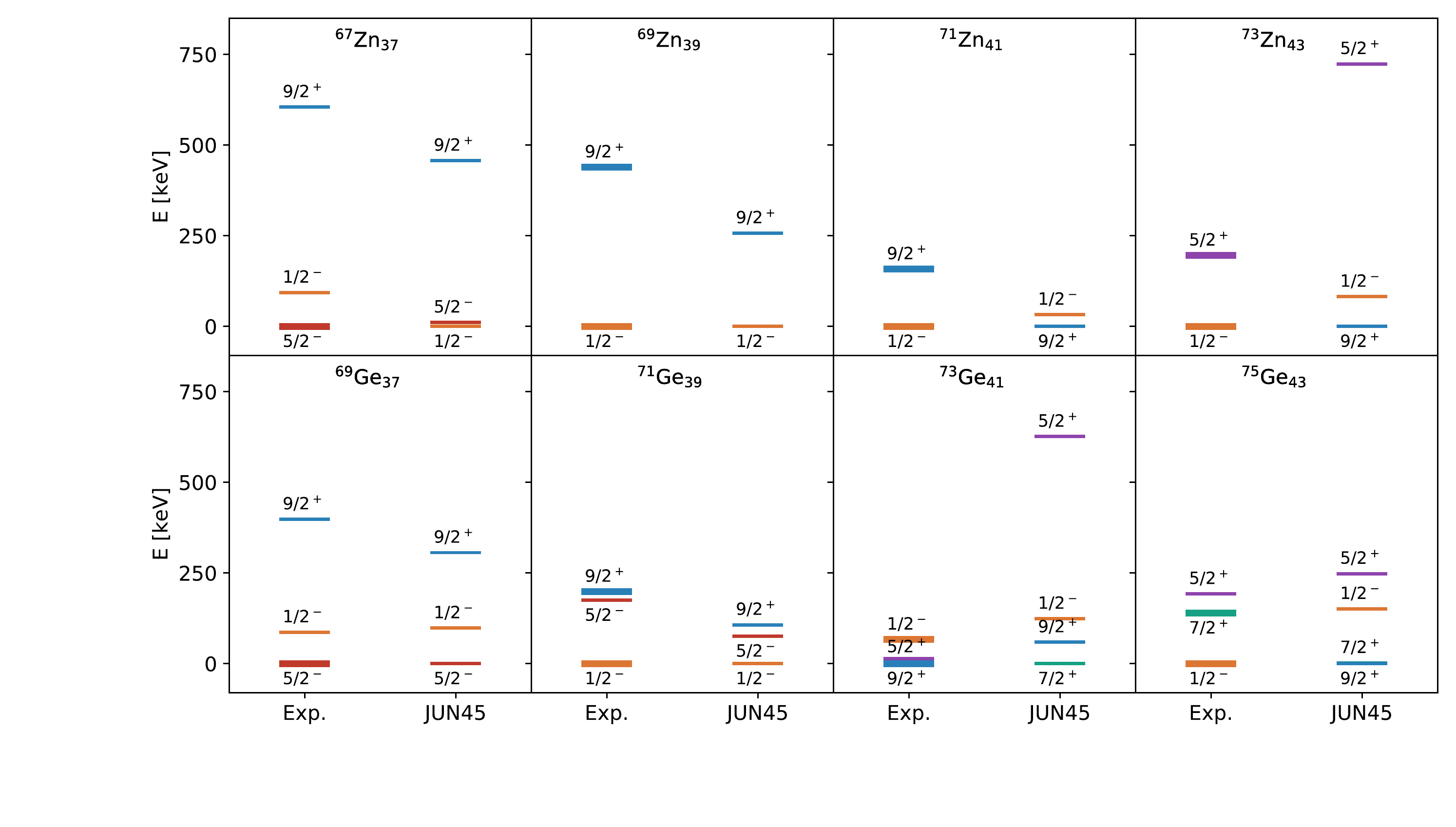}
\vspace{-3mm}
\caption{\label{fig:GeZnOddEnergy} Ground and low-lying excited (isomeric) states of $^{67-73}$Zn (top) and $^{69-75}$Ge (bottom) compared to results from JUN45 calculations. States with half-lives of at least 10 ns and a firm spin assignment are plotted. Thick lines represent states with half-lives longer than 1 ms.}
\vspace{-4mm}
\end{figure*}

The relevant proton and neutron shells in the nickel mass region are displayed in \mbox{Fig.~\ref{fig:shells}}. For the even-$Z$ elements (like nickel, zinc and germanium), with neutron numbers between $N = 28$ and $N = 50$, the neutrons are expected to dominate the ground state structure of the odd-$A$ isotopes in a naive single particle shell-model picture. By filling the $\nu 2p_{3/2}, \nu 1f_{5/2}, \nu 2p_{1/2}, \nu 1g_{9/2}$ orbitals, this would lead to ground-state (g.s.) spins of 3/2$^-$, 5/2$^-$, 1/2$^-$ and 9/2$^+$ for the odd-$A$ isotopes, respectively. Thus, for $^{69}$Ge and its isotones, $^{67}$Zn and $^{65}$Ni, with \mbox{$N = 37$}, the unpaired neutron is expected to occupy the $\nu f_{5/2}$ orbital, resulting in a g.s. spin of 5/2$^{-}$, which is consistent with the experimental assignment for each of these isotones~\cite{Browne2010, Junde2005, Nesaraja2014}. For isotones with \mbox{$N$ = 39}, a spin 1/2$^{-}$ is then expected and also experimentally confirmed for $^{67}$Ni, $^{69}$Zn and $^{71}$Ge~\cite{Junde2005, Nesaraja2014, Abusaleem2011}. So, below $N = 40$, the single-particle (SP) shell-model filling seems to be respected in germanium, despite having four valence protons, which could induce significant correlations and deformation. Moving to $N = 41$ and beyond, the filling of the $\nu g_{9/2}$ orbital is then beginning and the odd-$N$ isotopes are all expected to have  g.s. spin of 9/2$^{+}$. This is indeed the case for $^{73}$Ge and $^{69}$Ni but not for $^{71}$Zn~\cite{Nesaraja2014, Abusaleem2011, Singh2019}. This might suggest some stabilizing effect from a completely filled $\pi p_{3/2}$ orbital in $^{73}$Ge, and onset of correlations between protons in the open $\pi p_{3/2}$ orbital in $^{71}$Zn.

Recent studies have suggested that the structure of the zinc isotopes between $N = 40$ and $N = 50$~\cite{Wraith2017, Yang2018} presents some complexity with long-lived SP-like or deformed isomers occurring in each of the odd-$N$ isotopes.  For the germanium isotopes, deformation was proposed for the even-$A$ isotopes around $N=40$~\cite{Heyde2011, Heyde2016} but a systematic discussion of the nuclear moments of the odd-$N$ isotopes has not yet been done. We first investigate the nuclear structure around $N=40$ by looking at the low-lying energy levels of $^{67-73}$Zn and $^{69-75}$Ge isotopes, as summarized in \mbox{Fig.~\ref{fig:GeZnOddEnergy}}. Note that only the energy levels with a firm spin assignment and with measured half-lives longer than \mbox{10~ns} are plotted, as nuclear moments have been measured for most of them \cite{Mertzimekis2016}. These energy levels are also compared to large-scale shell-model calculations using the JUN45 effective interaction~\cite{Honma2009} in the $f_{5/2}pg_{9/2}$ model space starting from a $^{56}$Ni core, as illustrated in \mbox{Fig.~\ref{fig:shells}}.

While the shell-model reproduces well the level ordering in germanium and zinc isotopes below \mbox{$N = 40$}, it fails to reproduce the complex level structure in both isotopic chains beyond $N = 40$~\cite{Ayangeakaa2016, Sun2014, Toh2013, Yang2018}. This is apparent from the spins of the low lying states in $^{73, 75}$Ge and $^{73}$Zn, which may not be easily understood from the normal filling of the shell-model orbitals. In the $^{73}$Zn isotope, the isomeric character of the 5/2$^{+}$ state allowed a measurement of its nuclear moments, and an unambiguous assignment of its spins and parity to be made, which was debated before in literature~\cite{Wraith2017}. From the magnetic moment, it was clear that the unpaired neutrons mostly occupy the $g_{9/2}$ orbital, but the large quadrupole moment could be only explained by including proton and neutron excitations across $Z = 28$ and \mbox{$N = 50$}, leading to a triaxial shape~\cite{Yang2018}. Investigating the experimental electromagnetic properties of the low-lying states in the odd-$A$ germanium isotopes will thus help to get a better understanding of their structure, which will be discussed in the following sections.

\subsection{Magnetic Moments}
\begin{figure}[!t]
\includegraphics[width=0.5\textwidth]{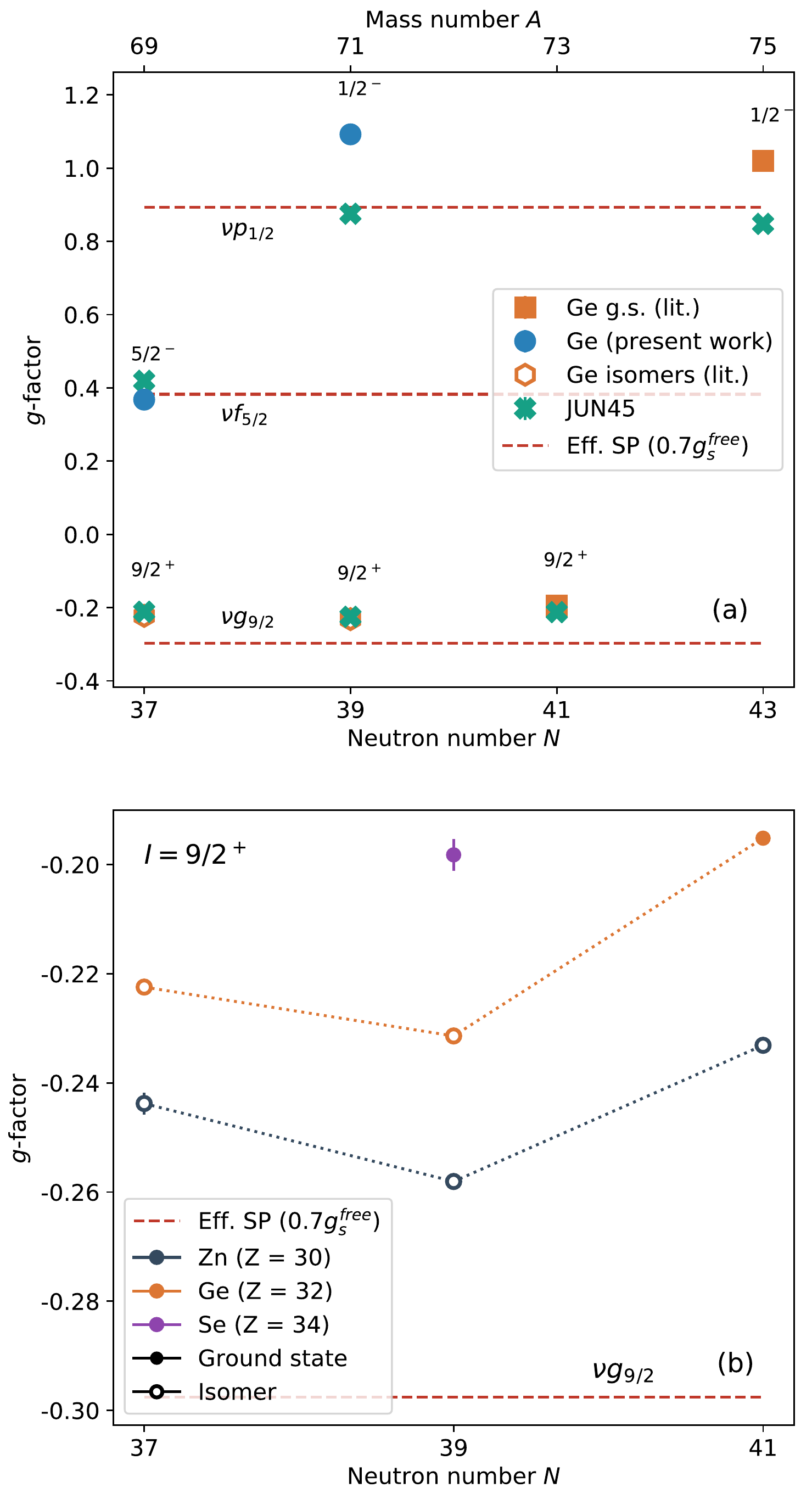}
\vspace{-10mm}
\caption{\label{fig:momentMu}(a) $g$-factor of ground and low-lying isomeric states of $^{67-75}$Ge, compared with the shell-model calculations using JUN45 interaction. (b) Experimental $g$-factors of ground and low-lying isomeric 9/2$^{+}$ states of the isotones of even-$Z$, zinc, germanium and selenium isotopes.}
\vspace{-5mm}
\end{figure}

The magnetic dipole moment, and more specifically the related $g$-factor ($g = \mu/I$), is an excellent probe of the orbital that is occupied by the unpaired particles (in this case neutrons). The available experimental $g$-factors of the ground and isomeric states of $^{69-75}$Ge are compared with the effective SP $g$-factors of relevant orbitals, as presented in \mbox{Fig.~\ref{fig:momentMu}(a)}. The effective SP $g$-factors for the orbitals of $\nu f_{5/2}$, $\nu p_{1/2}$, $\nu g_{9/2}$ are calculated using effective $g$-factors of $g_{s}^{\rm eff}$ = $0.7g_{s}^{\rm free}$ and $g_{l}^{\rm eff}=g_{l}^{\rm free}$, which are the typical values used in the region~\cite{Yang2016, Farooq-Smith2017}.

The newly measured magnetic moment (and $g$-factor) of the 5/2$^-$ g.s. of $^{69}$Ge ($N = 37$) is in excellent agreement with the $\nu f_{5/2}$ effective SP value, intuitively pointing to its simple shell-model character of an unpaired valence neutron in the $\nu f_{5/2}$ orbital. The available experimental $g$-factors for the states with spin 9/2$^+$ in $^{69,71,73}$Ge lie close to the effective SP $g$-factor of the $g_{9/2}$ orbital, as shown in \mbox{Fig.~\ref{fig:momentMu}(a)}. Again this suggests a configuration dominated by an unpaired neutron in the $\nu g_{9/2}$ orbital. More interestingly, the \mbox{$g$-factor} of the 9/2$^+$ state of $^{73}$Ge deviates more from the effective SP $g$-factor (enhanced in Fig. \ref{fig:momentMu}(b)), pointing to a possible onset of collective effects from $N=41$ onwards. As for the g.s. of $^{71,75}$Ge, both having a spin/parity of 1/2$^-$, the structure is likely dominated by an odd neutron in the $p_{1/2}$ orbital as the $g$-factors are close to the $\nu p_{1/2}$ SP value, yet deviate somewhat. Similar deviations have been observed for other 1/2$^-$ states, e.g. in the neighbouring zinc isotopes~\cite{Wraith2017}, as well as for heavier isotopes with spin 1/2$^-$ with a valence particle filling the $p_{1/2}$ orbital (e.g. Ag and In)~\cite{Eberz1987}.

These experimental $g$-factors are also compared with the shell-model calculations using the JUN45 interaction, as presented in \mbox{Fig.~\ref{fig:momentMu}(a)}. The calculations nicely reproduce the new experimental value of the 5/2$^-$ g.s. of $^{69}$Ge ($N = 37$) as well as the $g$-factors of the 9/2$^{+}$ states in $^{69-73}$Ge. However, the calculated wave functions for these states are found to be very fragmented. The calculated dominant configuration of the wave function for the 5/2$^-$ g.s. of $^{69}$Ge is about 26\%. For the 9/2$^{+}$ states of $^{69-73}$Ge, the main component of the wave function is, in all cases, less than 50\%. For all the isotopes, the leading configuration is the same as that concluded from the effective SP $g$-factors. Surprisingly, the calculated $g$-factors of the 1/2$^-$ g.s. of $^{71,75}$Ge are close to the SP values, so they also deviate from the experimental values. Further theoretical investigation is needed to understand this deviation for the 1/2$^-$ states.

Earlier studies have shown that the $N = 40$ subshell effect observed in the nickel and copper isotopes, quickly disappeared with more protons added above $Z = 28$, as described by various experimental observables: magnetic and quadrupole moments~\cite{Flanagan2009, DeGroote2017, Wraith2017, Cheal2010}, charge radii~\cite{Bissell2016, Xie2019, Procter2012}, nuclear masses~\cite{Rahaman2007, Guenaut2007}, $E(2^{+})$ excitation energies, and $B(E2)$ transition rates~\cite{Perru2006, Aoi2010, Chiara2011, Louchart2013}. This indicates that an increase in collectivity around and above $N$ = 40 is expected when going away from $Z$ = 28, as has been observed in the zinc and gallium isotopes~\cite{Cheal2010, Yang2018, Wraith2017, Xie2019, Perru2006, Aoi2010, Chiara2011, Louchart2013, Rahaman2007, Guenaut2007}. 

In order to have a systematic investigation of the structural evolution from single-particle nature to the collective effect, we compare the experimental magnetic moments of the 9/2$^{+}$ states for even-$Z$ nuclei (zinc, germanium and selenium) around $N = 40$, as shown in \mbox{Fig.~\ref{fig:momentMu}(b)}. The $g$-factors of all the 9/2$^{+}$ states in the region around $N$ = 40 are close to the effective SP $g$-factor, clarifying the leading configuration of an unpaired neutron in the $g_{9/2}$ orbital in their wave functions. However, a continuously increased deviation from the SP value is obvious from zinc, to germanium and selenium, giving a clear sign of the increased collectivity as more protons are added above $Z = 28$ closed shell. This is further confirmed by comparing the contribution of the main configuration of wave functions (single neutron in $g_{9/2}$ orbital), which is about $\sim$35\% for $^{69}$Zn and $\sim$14\% for $^{71}$Ge.

\begin{figure}[!t]
\includegraphics[width=0.5\textwidth]{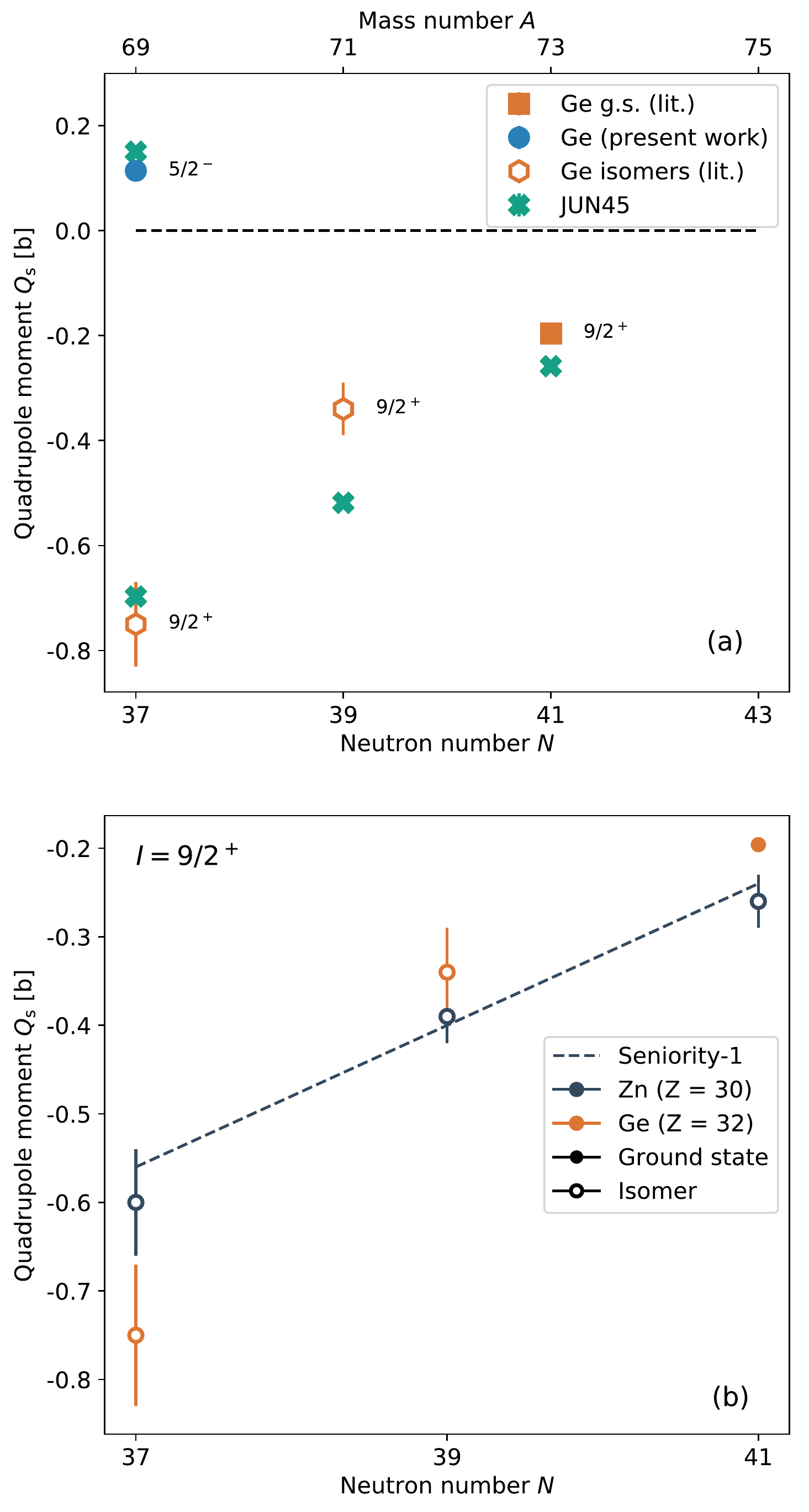}
\vspace{-5mm}
\caption{\label{fig:momentQs}(a) Quadrupole moments of ground and low-lying isomeric states of $^{67-75}$Ge, compared with the shell-model calculations using JUN45 interaction using effective charges $e_{p}^{\rm eff}$ = $1.5e_{p}$ and $e_{n}^{\rm eff}=1.1e_{n}$. (b) Experimental quadrupole moments of ground and low-lying isomeric 9/2$^{+}$ states of even-$Z$ zinc, germanium and selenium.}
\vspace{-5mm}
\end{figure}

\subsection{Quadrupole moments}
The known experimental quadrupole moments of ground and isomeric states in $^{69-75}$Ge are presented in \mbox{Fig.~\ref{fig:momentQs}(a)}. The new value of the quadrupole moment of the 5/2$^-$ g.s. in $^{69}$Ge is well reproduced by the shell-model calculation. The same agreement can be found for the 9/2$^{+}$ states of $^{71-73}$Ge. It is known from the above-discussed magnetic moments that the main configuration for these 9/2$^+$ states comes from the unpaired valence neutron in the $\nu g_{9/2}$ orbital. 

In \mbox{Fig.~\ref{fig:momentQs}(a)}, we see a linear trend in the quadrupole moments of these 9/2$^+$ states, when more neutrons fill the $\nu g_{9/2}$ orbital. A similar trend has been observed in the neighboring zinc isotopes~\cite{Wraith2017, Yang2018}, which are compared to those of germanium in \mbox{Fig.~\ref{fig:momentQs}(b)}. In the case of the zinc isotopic chain, the quadrupole moments of the 9/2$^+$ states in $^{69}$Zn ($N=39$) and $^{79}$Zn ($N=49$) reflect a relatively pure configuration of single neutron particle and single neutron hole in the $\nu g_{9/2}$ orbital, respectively~\cite{Wraith2017, Yang2018}, from which we estimated the SP quadrupole moment of $\nu g_{9/2}$ orbital (see Ref.~\cite{Wraith2017} for more details). A linear trend for quadrupole moments with increased neutron number for a seniority-1 $(\nu g_{9/2})^n$ configuration is then calculated, as shown with the black dash line in \mbox{Fig.~\ref{fig:momentQs}(b)}. 

For the Ge isotopes, with only three measured quadrupole moments, it is difficult to make a firm conclusion. It appears that the slope of the curve is somewhat steeper than for the Zn isotopes, which would point to some enhanced deformation (correlations) for these isotopes with four protons outside the $Z=28$ shell. However, more data on neutron-rich isotopes are needed to make a firm conclusion.

\section{Summary}
The hfs of germanium atoms was measured for the first time in the $4s^2 4p^2 \, ^3P_1 \rightarrow 4s^2 4p 5s \, ^3P_1^o$ atomic transition, taking advantage of the development of the laser frequency-mixing technique. A clear discrepancy with literature magnetic and electric hyperfine parameters is observed for $^{69}$Ge. The systematic analysis of the hfs of $^{69,71,73}$Ge isotopes obtained in this work requires a revision of the hyperfine constants for $^{69}$Ge, resulting in new magnetic and quadrupole moments. State-of-the-art atomic relativistic Fock-Space Coupled-Cluster calculations were performed for the hyperfine fields in three atomic fine structure levels, allowing to determine the quadrupole moment of $^{73}$Ge to a precision of 2\%, in agreement with the precision value obtained from molecular theory calculations and experiments. 

The available experimental nuclear moments of ground and isomeric states of $^{69-75}$Ge around $N = 40$ are interpreted through a comparison to large-scale shell-model calculations in the $f_{5/2}pg_{9/2}$ model space. A comparison of the $g$-factors with the effective SP \mbox{$g$-factors} reveals the nature of the orbital occupied by the unpaired neutrons. Yet, the calculated wave functions reveal a very mixed configuration. Through a systematic comparison of the nuclear moments of germanium isotopes with its isotones (zinc and selenium) around $N = 40$, an increase in collectivity is observed, when protons are added to $Z=28$, reflecting a sign of structural change from single particle to deformation when moving away from $Z = 28$.

To shed more light on the structural evolution in the region, further study of germanium isotopes up to $N = 50$ and even beyond is necessary, with the well-established laser systems and atomic transition studied in this work. This will allow us to have a better understanding of the structure changes in the nickel region, particularly, the gradual emergence of collectivity. \\

\begin{acknowledgments}
We acknowledge the support of the ISOLDE collaboration and technical teams. We would like to acknowledge the Center for Information Technology of the University of Groningen for their support and for providing access to the Peregrine high-performance computing cluster. This work was supported by the National Key R\&D Program of China (No. 2018YFA0404403), the National Natural Science Foundation of China (No:11875073, U1967201); the BriX Research Program No. P7/12, FWO-Vlaanderen (Belgium), GOA 15/010 from KU Leuven; the UK Science and Technology Facilities Council grants ST/L005794/1 and ST/P004598/1; the NSF grant PHY-1068217, the BMBF Contracts No 05P18RDCIA, the Max-Planck Society, the Helmholtz International Center for FAIR (HIC for FAIR); the EU Horizon2020 research and innovation programme through ENSAR2 (No. 654002);
\end{acknowledgments}

\bibliography{Ge_moments_biblio} 

\end{document}